\documentclass[aps,pra,superscriptaddress,floatfix,showpacs,10pt,twocolumn]{revtex4}
\usepackage{amsmath}
\usepackage{graphicx}
\usepackage{amsfonts}
\usepackage{amssymb}

\usepackage[T1]{fontenc}
\usepackage[latin2]{inputenc}
\usepackage{latexsym}
\usepackage{bbm}

\usepackage{color}

\begin{document}
\title{Conditions for the freezing phenomena of geometric measure of quantum discord for arbitrary two-qubit X-states under non-dissipative dephasing noises}
\author{Wei Song}
 \affiliation{School of Electronic and Information Engineering, Hefei Normal University, Hefei 230061,
China}

\author{Zhuo-Liang Cao}
\affiliation{School of Electronic and Information Engineering, Hefei
Normal University, Hefei 230061, China}

\date{\today}


\begin{abstract}
We study the dynamics of geometric measure of quantum discord (GMQD)
under the influences of two local phase damping noises. Consider the
two qubits initially in arbitrary X-states, we find the necessary
and sufficient conditions for which GMQD is unaffected for a finite
period. It is further shown that such results also hold for the
non-Markovian dephasing process.
\end{abstract}
\maketitle

Quantum discord has received a lot of attentions due to its
potential to serve as an important resource in the deterministic
quantum computation with one pure
qubit(DQC1)\cite{Datta:2008,Modi:2011}. It indicates that
unentangled states may process quantumness which can be exploited in
certain types of quantum information processing tasks. There are
several versions of quantum discord. The first definition of quantum
discord is introduced by Ollivier and Zurek \cite{Ollivier:2001}and,
independently, by Henderson and Vedral \cite{Henderson:2001}. The
quantum discord of a composite system AB is defined by $D_A  \equiv
\mathop {\min }\limits_{\left\{ {E_i^A } \right\}} \sum\limits_i
{p_i } S\left( {\rho _{B|i} } \right) + S\left( {\rho _A } \right) -
S\left( {\rho _{AB} } \right)$, where $S\left( {\rho _{AB} } \right)
= Tr\left( {\rho _{AB} \log _2 \rho _{AB} } \right)$ is the von
Neumann entropy and the minimum is taken over all positive operator
valued measures(POVMs) $\left\{ {E_i^A } \right\}$ on the subsystem
$A$ with $p_i  = Tr\left( {E_i^A \rho _{AB} } \right)$ being the
probability of the $i$-th outcome and $\rho _{B|i}  = {{Tr_A \left(
{E_i^A \rho _{AB} } \right)} \mathord{\left/
 {\vphantom {{Tr_A \left( {E_i^A \rho _{AB} } \right)} {p_i }}} \right.
 \kern-\nulldelimiterspace} {p_i }}$ being the conditional state of
 subsystem $B$. The minimum can also be taken over the von Neumann
 measurements. Because of the minimization taken over all possible POVMs, or von Neumann
measurements, it is generally difficult to get analytical results.
In order to overcome this difficulty geometric measure of quantum
discord (GMQD) has been introduced by Dakic \emph{et al}
\cite{Dakic:2010}. GMQD is defined by $ D_A^g = \mathop {\min
}\limits_{\chi  \in \Omega _0 } \left\| {\rho - \chi } \right\|^2 $,
where $ \Omega _0 $ denotes the set of zero-discord states and $
\left\| {X - Y} \right\|^2  = Tr\left( {X - Y} \right)^2 $ is the
square norm in the Hilbert-Schmidt space. The subscript $A$ denotes
that the measurement is taken on the system $A$. Especially, Dakic
\emph{et al} \cite{Dakic:2012} showed that the GMQD is related to
the fidelity of remote state preparation which provides an
operational meaning to GMQD.

Because of the unavoidable interaction between a quantum system and
its environment, another interesting area of investigation of the
discord is its behaviour under different noisy environments
\cite{Werlang:2009b,Maziero:2009,Wang:2010,Mazzola:2010,Li:2011,Fanchini:2010}.
It has been shown that quantum discord is more robust than
entanglement for both Markovian and non-Markovian dissipative
processes. In particular, the discord of a sort of Bell-diagonal
states subject to a phase damping noises was shown to exhibit a
freezing phenomenon, \cite{Mazzola:2010}i.e., the quantum discord
can be completely unaffected by decoherence for a finite period of
time. Lang and Caves \cite{Lang:2010} provide a complete geometric
picture for the frozen-discord phenomenon. The freezing phenomena
have been found to be a robust feature of a class of models in the
presence of non-dissipative decoherence
\cite{Li:2011,Mazzola:2011,Karpat:2011}. The model of Mazzola
\emph{et al} has also been extended to local non-Markovian case
\cite{Mazzola:2011}. The experimental demonstration of this
phenomenon have been reported by Xu \emph{et al.} \cite{Xu:2010a}
with optical systems and Auccaise \emph{et al} \cite{Auccaise:2011}
using NMR. However, previous works mainly focused on the quantum
discord based on Ref. \cite{Ollivier:2001,Henderson:2001} and thus
were restricted to some special cases such as Bell-diagonal states.
It has been shown that the quantum discord and GMQD do not
necessarily imply the same ordering of two-qubit X-states
\cite{Altintas:2010,Xu:2009,Okrasa:2011,Yeo:2010}. More recently,
Girolami and Adesso \cite{Girolami:2011} and independently Batle
\emph{et al.} \cite{Batle:2011} provided a numerical evidence, from
which one can infer that there exist other states violating the
states ordering with quantum discords. It is thus worth
investigating the freezing phenomena with GMQD. In this paper we
consider the two qubits initially in arbitrary X-states in the
presence of local dephasing noises, and we derive the conditions for
the freezing phenomenon in terms of GMQD. We also consider the
non-Markovian dephasing case and find that our results also hold for
the non-Markovian dephasing process. The fact that, under certain
conditions, GMQD remains unchanged under the phase damping noise for
some period of time may have potential applications in future
quantum information tasks.

Before describing the main results of this paper, let us briefly
review the method to compute GMQD. An arbitrary two-qubit state can
be written in Bloch representation:

\begin{eqnarray}
 \rho  = \frac{1} {4}\left[ {I \otimes I + \sum\limits_i^3 {\left(
{x_i \sigma _i  \otimes I + y_i I \otimes \sigma _i } \right) +
\sum\limits_{i,j = 1}^3 {R_{ij} \sigma _i  \otimes \sigma _j } } }
\right]
\end{eqnarray}

\noindent where $ x_i  = Tr\rho \left( {\sigma _i  \otimes I}
\right),y_i  = Tr\rho \left( {I \otimes \sigma _i } \right) $ are
components of the local Bloch vectors, $ \sigma _i ,i \in \left\{
{1,2,3} \right\} $ are the three Pauli matrices, and $ R_{ij}  =
Tr\rho \left( {\sigma _i  \otimes \sigma _j } \right) $ are
components of the correlation tensor. For two-qubit case, the
zero-discord state is of the form $ \chi  = p_1 \left| {\psi _1 }
\right\rangle \left\langle {\psi _1 } \right| \otimes \rho _1  + p_2
\left| {\psi _2 } \right\rangle \left\langle {\psi _2 } \right|
\otimes \rho _2$, where $ \left\{ {\left| {\psi _1 } \right\rangle
,\left| {\psi _2 } \right\rangle } \right\} $ is a single-qubit
orthonormal basis. Then a analytic expression of the GMQD is given
by \cite{Dakic:2010}

\begin{eqnarray}
D_A^g \left( \rho  \right) = \frac{1} {4}\left( {\left\| x
\right\|^2  + \left\| R \right\|^2  - k_{\max } } \right)
\end{eqnarray}

\noindent where $ x = \left( {x_1 ,x_2 ,x_3 } \right)^T $ and $
k_{\max } $ is the largest eigenvalue of matrix $ K = xx^T  + RR^T
$. By introducing a matric $\mathcal{R}$ defined by

\begin{equation}
\label{eq5} \mathcal{R}  = \left( {{\begin{array}{*{20}c}
1 & {y^T }  \hfill \\
x & R \hfill \\
\end{array} }} \right)
\end{equation}
\noindent and $ 3 \times 4$ matric $\mathcal{ {R'} }$ through
deleting the first row of $\mathcal{R}$, the analytical expression
of GMQD can be further rewritten as \cite{Lu:2010}

\begin{eqnarray}
D_A^g \left( \rho  \right) = \frac{1} {4}\left[ {\left(
{\sum\limits_k {\lambda _{_k }^2 } } \right) - \mathop {\max
}\limits_k \lambda _{_k }^2 } \right]
\end{eqnarray}
\noindent where $ {\lambda _k } $ is the singular values of
$\mathcal{R'}$. In the following discussions we use Eq.(4) to
calculate GMQD. Suppose the initial state is prepared in arbitrary
two-qubit X-states with the general form

\begin{equation}
\label{eq5} \rho_X   = \left( {{\begin{array}{*{20}c}
\rho_{11} & 0 & 0  & \rho_{14} \hfill \\
0 &\rho_{22}& \rho_{23}  & 0 \hfill \\
0 & \rho_{32} & \rho_{33} & 0 \hfill \\
\rho_{41} & 0 & 0  & \rho_{44} \hfill \\

\end{array} }} \right)
\end{equation}

\noindent where we have chosen the computational basis $\left\{
{\left| {00} \right\rangle ,\left| {01} \right\rangle ,\left| {10}
\right\rangle ,\left| {11} \right\rangle } \right\}$. Eq.(5) is a
7-real parameter state with three real parameters along the main
diagonal and two complex parameters at off-diagonal positions. A
remarkable aspect of the X-states is that the initial X structure is
preserved during the decoherence process, thus, it is convenient to
get analytical results. To get analytical results generally, we use
GMQD to quantify the quantum correlation contained in the X-states
and our results are summarized as the following theorem:

\noindent \textbf{ Theorem}: Consider arbitrary  X-states defined in
Eq.(5), with each qubit being subject to the local phase damping
noises. The necessary and sufficient conditions for the freezing
phenomena of GMQD are:

\noindent
\begin{eqnarray}
\left| {\rho _{14} } \right| &=& \left| {\rho _{23} } \right|,\nonumber\\
8\left| {\rho _{14} \rho _{23} } \right| &>& \left( {\rho _{11} -
\rho _{33} } \right)^2  + \left( {\rho _{22}  - \rho _{44} }
\right)^2
\end{eqnarray}

\noindent \textbf{Proof.} First we notice that the X-states defined
in Eq.(5) can be rewritten in the Bloch representation with the
correlation matrix $R$ given by
\begin{widetext}
\begin{equation}
\label{eq5} R  = \left( {{\begin{array}{*{20}c}
\rho_{14}+\rho_{23}+\rho_{32}+\rho_{41} & i(\rho_{14}-\rho_{23}+\rho_{32}-\rho_{41}) & 0   \hfill \\
i(\rho_{14}+\rho_{23}-\rho_{32}-\rho_{41}) &-\rho_{14}+\rho_{23}+\rho_{32}-\rho_{41}& 0  \hfill \\
0 & 0 &\rho_{11}-\rho_{22}-\rho_{33}+\rho_{44}   \hfill \\

\end{array} }} \right)
\end{equation}
\end{widetext}

\noindent and $ x = \left( {0,0,\rho _{11}  + \rho _{22}  - \rho
_{33}  - \rho _{44} } \right)^T $. Thus the GMQD can be calculated
according to Eq.(4) which is given by $ D_{_A }^g \left( \rho
\right) = \frac{1} {4}\left( {\lambda _1^2  + \lambda _2^2  +
\lambda _3^2  - \max \left\{ {\lambda _1^2 ,\lambda _2^2 ,\lambda
_3^2 } \right\}} \right) $, where

\begin{eqnarray}
\lambda _1^2  & =& 4\left( {\left| {\rho _{14} } \right|^2  + \left|
{\rho _{23} } \right|^2 } \right) + 8\left| {\rho _{14} \rho _{23} }
\right|
\nonumber\\
\lambda _2^2  & =& 4\left( {\left| {\rho _{14} } \right|^2  + \left|
{\rho _{23} } \right|^2 } \right) - 8\left| {\rho _{14} \rho _{23} }
\right|
\nonumber\\
\lambda _3^2  & =& 2\left[ {\left( {\rho _{11}  - \rho _{33} }
\right)^2  + \left( {\rho _{22}  - \rho _{44} } \right)^2 } \right]
\end{eqnarray}

We suppose the X-states is subjected to two local Markovian phase
damping noises formulated via its Kraus representation
as\cite{Neilsen:2000,Horst:2013} $ \Im
 \left( \rho \right) = E_1 \rho E_1^\dag   + E_2 \rho
E_2^\dag $, with $E_1  = \sqrt {1 - p} \left| 1 \right\rangle
\left\langle 1 \right|  $, and $ E_2  = \left| 0 \right\rangle
\left\langle 0 \right| + \sqrt p \left| 1 \right\rangle \left\langle
1 \right| $. The parameter $p$ ranges from 0 to 1. The phase damping
noises will affect the off-diagonal elements while the diagonal
elements remain unchanged. In order to exhibit a freezing
phenomenon, the GMQD of the X-states should not contain the
parameter $p$ after subjected to the phase damping nosies.  If
$\left| {\rho _{14} \rho _{23} } \right| \ne 0 $, then $ {\lambda
_1^2  > \lambda _2^2 } $. According to the formula of GMQD, it is
directly to see that the condition of GMQD remain unaffected is
given by $ {\lambda _2^2  = 0} $ and $ {\lambda _1^2  > \lambda _3^2
} $. Combined with these conditions, the theorem is proved.
$\hfill\blacksquare$

In order to show the application of our theorem we consider a
subclass of two qubit X-states given by

$ \chi = \frac{1} {4}\left[ {I \otimes I + \textbf{r} \cdot \sigma
\otimes I + I \otimes \textbf{s} \cdot \sigma  + \sum\limits_{i =
1}^3 {c_i \sigma _i \otimes \sigma _i } } \right] $

\noindent where we choose the Bloch vectors in $z$ direction with $
\textbf{r} = \left( {0,0,r} \right)$,$ \textbf{s} = \left( {0,0,s}
\right) $. According to Eq.(6), the condition becomes $ {c_1^2  =
0,c_2^2
> r^2  + c_3^2 } $ or ${c_2^2  = 0,c_1^2  > r^2 + c_3^2 } $ which
reduces to the result presented in Ref.\cite{Song:2011}. For fixed
parameters $r$ and $s$, the above inequalities become a
two-parameter set, whose geometry can be depicted. Combined with the
positivity condition of the eigenvalues of the density matrix $\chi$
we can depict the possible region of the mixed state. In Fig.1 we
plot the physical region with different $r$ and $s$, respectively.
If $r = s = 0$, the mixed state reduces to the two-qubit
Bell-diagonal states.

\begin{figure}[ptb]
\includegraphics[scale=0.70,angle=0]{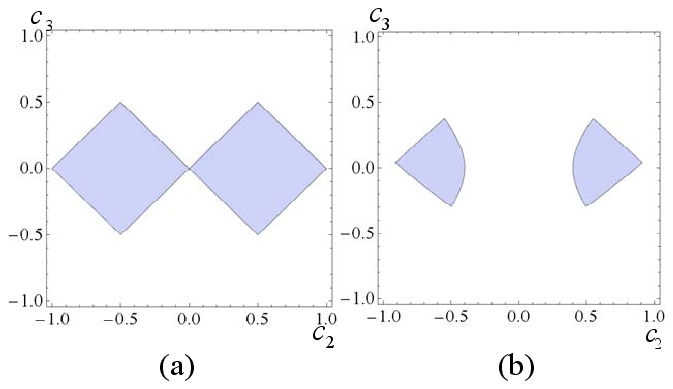}\caption{(Color online). The geometry of the set of valid
states satisfying Eq.(6) with different $r$ and $s$, repectively.
(a)$r$=$s$=0,(b)$r$=0.4,$s$=0.1}
\label{fig1}%
\end{figure}

\begin{figure}[ptb]
\includegraphics[scale=0.60,angle=0]{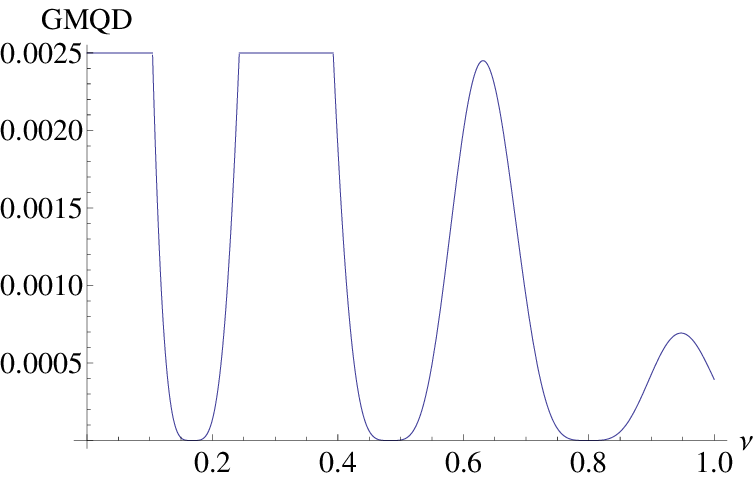}\caption{(Color online). The dynamics of GMQD of Bell-diagonal states interacting with non-Markovian dephasing channels,
where we have chosen $c_1  = 0,c_2  = 0.35,c_3  = 0.1$, $\nu  =
\frac{t} {{2\tau }}$ with $\tau  = 2.5s$, $a $= 1${s^{ - 1} }$.}
\label{fig1}%
\end{figure}

Further we find that our result also holds for the non-Markovian
dephasing process. The reason is that the diagonal elements of the
X-states remain unchanged in the non-Markovian case, thus the proof
does not altered. For concreteness, let us consider Bell-diagonal
state $\rho _{AB}  = \frac{1} {4}\left( {{\rm I}_{AB}  +
\sum\limits_{i = 1}^3 {c_i \sigma _i^A \sigma _i^B } } \right)$
subject to the non-Markovian model presented by Daffer \emph{et al}
\cite{Daffer:2004}, where ${\sigma _i^{A\left( B \right)} }$ denote
the Pauli operators in direction $i$ acting on A(B). It is shown
that the map has a Kraus decomposition $\Phi _t \left( \rho \right)
= \sum\limits_n {A_n^\dag  } \rho A_n $, with $A_i  = \sqrt {{{1 -
\Lambda \left( \nu  \right)} \mathord{\left/
 {\vphantom {{1 - \Lambda \left( \nu  \right)} 2}} \right.
 \kern-\nulldelimiterspace} 2}} \sigma _i ,A_j  = 0,A_k  = 0,A_4  = \sqrt {{{1 + \Lambda \left( \nu  \right)} \mathord{\left/
 {\vphantom {{1 + \Lambda \left( \nu  \right)} 2}} \right.
 \kern-\nulldelimiterspace} 2}} I$, where $\Lambda \left( \nu  \right) = e^{ - \nu } \left[ {\cos \left( {\mu \nu } \right) + \frac{{\sin \left( {\mu \nu } \right)}}
{\mu }} \right],\mu  = \sqrt {\left( {4a\tau } \right)^2  - 1} ,\nu
= \frac{t} {{2\tau }}$ is the dimensionless time, $i=1,2,3$ denote
the direction of noise, and $j$ and $k$ denotes the directions in
which there is no noise. By changing the direction of the noise we
can get a colored noise bit flip, bit-phase flip or phase flip
channels, respectively. This process is non-Markovian according to
the measure of non-Markovianity of Ref.\cite{Breuer:2009}. Using the
above Kraus decomposition we obtain the parameters of $c_i$ of the
Bell-diagonal states evolve as: $c_3 \left( t \right) = c_3 ,c_1
\left( t \right) = c_1 \Lambda \left( \nu \right)^2 ,c_2 \left( t
\right) = c_2 \Lambda \left( \nu \right)^2 $, where we have chosen
the phase flip channel(which means $i=3$) case and similar results
can be found for bit flip and bit phase flip channels. For fixed
$c_i $ we plot the dynamic behavior of GMQD in Fig.2. In this case,
it is shown that the frozen phenomenon appear two times in contrast
to Markovian case where the frozen phenomenon appears only one time
\cite{Mazzola:2010}.

In summary, we have investigated the dynamics of GMQD for X-states
under the local dephasing noises. We derived the necessary and
sufficient conditions for freezing phenomenon in terms of GMQD. We
also considered the non-Markovian dephasing case and that our
results can also be applied to the non-Markovian dephasing process.
Compared with the work in Ref.\cite{You:2012}, our result provides a
different perspective from the geometric measure. In particular, it
holds for the more general case which does not restricted to the
Bell-diagonal states. An open question is whether there exist other
classes of mixed states exhibiting the frozen phenomenon in terms of
GMQD. The future study involves exploring potential applications to
exploit the class of initial states exhibiting frozen phenomenon in
future quantum information tasks without any disturbance from the
noisy environment.

This work was supported by National Natural Science Foundation of
China under Grant No.10905024, No.11374085, No.11204061,
No.61073048, the Key Project of Chinese Ministry of Education under
Grant No.211080, and the Key Program of the Education Department of
Anhui Province under Grant No.KJ2011A243, No.KJ2012A244.

\end{document}